\documentclass[twocolumn]{aastex63}

\received{}
\revised{}
\accepted{}

\submitjournal{ApJ}

\shorttitle{Solar astrometric jitter}
\shortauthors{Shapiro et al.}

\graphicspath{{./}{figures/}}

\usepackage{amsmath}
\begin{document}

\title{Predictions of astrometric jitter for Sun-like stars. I. The model and its application to the Sun as seen from the ecliptic}

\author{Alexander I. Shapiro}
\email{shapiroa@mps.mpg.de}
\affiliation{Max-Planck-Institut fur Sonnensystemforschung, Goettingen, Germany}

\author{Sami K. Solanki}
\email{solanki@mps.mpg.de}
\affiliation{Max-Planck-Institut fur Sonnensystemforschung, Goettingen, Germany}
\affiliation{School of Space Research, Kyung Hee University, Yongin, Gyeonggi 446-701, Korea}

\author{Natalie A. Krivova}
\affiliation{Max-Planck-Institut fur Sonnensystemforschung, Goettingen, Germany}

\begin{abstract}
The advent of Gaia, capable of measuring stellar wobbles caused by orbiting planets, raised an interest to the astrometric detection of exoplanets. Another source of such wobbles (often also called jitter) is stellar magnetic activity. A quantitative assessment of the stellar astrometric jitter is important for a more reliable astrometric detection and characterisation of exoplanets. We calculate the displacement of the solar photocentre due to the magnetic activity for an almost 16-year period (February 2, 1999--August 1, 2014). We also investigate how the displacement depends on the spectral passband chosen for observations, including the wavelength range to be covered by the upcoming Small-JASMINE mission of JAXA.  This is done by extending the SATIRE-S model for solar irradiance variability to calculating the displacement of the solar photocentre caused by the magnetic features on the surface of the Sun. We found that the peak to peak amplitude of the solar photocentre displacement would reach 0.5 $\mu$as if the Sun were located 10 pc away from the observer and observed in the Gaia G filter. This is by far too small to be detected by the Gaia mission. However, the Sun is a relatively inactive star so that one can expect significantly larger signals for younger, and, consequently, more active stars. The model developed in this study can be combined with the simulations of emergence and surface transport of magnetic flux which have recently became available to  model the astrometric jitter over the broad range of magnetic activities.  
\end{abstract}

\section{Introduction}\label{sect:intro}
The discovery of an exoplanet rotating around a main-sequence star \citep{exo} initiated the new, highly dynamical field of exoplanetary science. As of today, more than 4000  planets have been confirmed\footnote{\url{http://exoplanet.eu/catalog}}. Presently the main techniques for discovering exoplanets are transit photometry and radial velocity  measurements. In particular, there have been no reliable astrometric detections until now. However, it is expected that the situation will soon change due to the advent of the ultra-precise astrometry. For example, it has been estimated \citep{Perryman2014} that  astrometric measurements by the Gaia mission \citep{gaia} would lead to a discovery of more than 20000 exoplanets. The main idea behind the astrometric approach is to detect the displacement of the stellar photocentre caused by the rotation of a star around star-planet barycentre. Interestingly, apart from the exoplanets the stellar photocentre can also be displaced by the magnetic activity of a star, since bright and dark magnetic features on a stellar surface affect the position of the photocentre \citep[see, e.g.,][]{Lanza2008}. {We note that  the activity induced astrometric and photometric signals have common origin. Dark spot on the stellar disk not only reduces stellar brightness but also repeals the stellar photocentre, e.g. spot at the western part of the stellar disk will shift the photocentre to the eastern part. Likewise, bright facular region increases stellar brightness and attracts the photocentre.} Consequently, similarly to the case of photometric and radial velocity measurements the intrinsic astrometric jitter from the host stars might become a hurdle for detecting and characterising exoplanets with astrometry.

A significant effort \citep[see, e.g.][ and references therein]{Eriksson2007,Catanzarite2008} has been invested in modelling such an astrometric jitter in anticipation of the planned (but never realised) Space Interferometry Mission.  In particular, \cite{Makarov2009} developed an analytical model attributing photometric, astrometric, and radial velocity (RV) jitter to a single spot on the stellar surface. They estimated that the solar jitter can reach a value of 1.5 $\mu$AU. \cite{Makarov2010} used Mount Wilson magnetograms and intensity images to create disk maps of solar bolometric surface intensity and to directly calculate the position of the solar photocentre. They found that largest deviation reaches 2.6 $\mu$AU and the standard deviation of the solar photocentre in the year 2000 was 0.91 $\mu$AU. A similar result was later obtained by \cite{Lagrange2011}, who used a more comprehensive model, developed by \cite{Meunier2010}, and MDI/SOHO magnetogramms to conclude that the solar astrometric jitter is most of the time smaller than 2 $\mu$AU. Recently, \cite{Meunier2019} outlined how to extend these calculations to other Sun-like stars. 

Lately, the motivation to model astrometric jitter has been boosted by data from the Gaia mission \citep[e.g. astrometric data from the first 22 months of observations have been recently released, see][]{Gaia_astrom}. For example, \cite{Morris2018} constructed a model attributing solar and stellar astrometric jitter to dark spots (i.e. neglecting contributions from bright faculae) and found that the precision of Gaia should be sufficient to detect astrometric jitter in the nearest active stars.

Concurrently with studies of the stellar jitter  a number of models of solar irradiance variability have been created \citep[see, e.g.,][for reviews]{TOSCA2013, MPS_AA}. The main motivation for the development of such models came from a suspected link between the solar irradiance variability and the natural climate  change \citep[see, e.g., review by][and references therein]{grayetal2010}. 
Later, some of these solar models, in particular the SATIRE \citep[Spectral And Total Irradiance Reconstruction][]{fliggeetal2000,krivovaetal2003} model, were successfully applied to the analysis and explanation of the photometric data  from Sun-like stars 
\citep[e.g.][]{Shapiro2014_stars, witzkeetal2018, Karoff2018, Timo2018}.

In this paper, we take a first step in applying the solar paradigm to calculating astrometric jitter from Sun-like stars. To that end, we extend SATIRE model  to calculating the astrometric jitter produced by the Sun. In particular, we have calculated solar astrometric jitter as it would be seen by the Gaia and Small-JASMINE
mission, which is a JAXA space mission for monitoring the distances and motions of stars in the near-infrared \citep[planned to be launched in 2024, see][]{JASMINE}. {However, before utilising the full power of the SATIRE approach we present a simple and more straightforward estimate of the astrometric jitter in Sect.~\ref{sect:estimate}}.


{
\section{Single-spot estimate}\label{sect:estimate}
The main goal of this section is to provide a very simplified estimate of the amplitude of the astrometric jitter, which can be expected for Sun-like stars before indulging in developing a much more realistic (and also sophisticated) model. Namely, we connect here the amplitudes of the stellar astrometric jitter and brightness variability assuming that both phenomena are attributed to the periodic transits of a not-evolving single spot over the visible stellar disk as the star rotates. 
We note that such an assumption is not expected to accurately represent the solar case. Indeed, the variability of the Sun is brought about by multiple spot and facular features whose distribution on the solar surface is rather intricate \citep[see, e.g.,][]{Sami_B}. A single-spot estimate might also be oversimplified for calculating the astrometric jitter of   young cool stars since it was suggested that some of them have a large number of spots on their surfaces \citep{JacksonandJeffries2012}. At the same time \cite{Emre2020} recently proposed that the variability of highly variable stars with near-solar fundamental parameters and rotation periods \citep[see][]{Jinghua2020, Timo2020} could be explained by the strong degree of nesting of magnetic features on their surfaces. Such highly-nested distributions of magnetic features could be much better approximated by a single-spot model than that of the Sun \citep[see, e.g. Fig.~2 from][]{Emre2020}.  

We consider a case of a star observed from its equatorial plane and a spot located in the stellar equator. Let us further assume that brightness contrast of the spot with respect to the quiet stellar regions does not depend on the position of the spot on the stellar disk \citep[which is a reasonable simplification, see e.g., Fig.~4 from][]{Shapiro2014_stars}. In such a case the relative drop of the stellar brightness due to the spot located on the visible disk can be written as $\Delta \cos\phi$, where $\Delta$ is a drop of the brightness when the spot is located in the visible disk centre and $\phi$ is the heliocentric angle of the spot (so that $\cos\phi$ represents foreshortening, i.e. it is equal to 1 for the spot located in the centre of the visible disk and 0 for the spot located at the limb). At the same time the shift of the photocentre can be written as   ${\cal R} \cdot \Delta \cos\phi \sin\phi$, where ${\cal R}$ is the visible angular radius of the star. Consequently, the maximum shift of the photocentre from the stellar disk centre corresponds to $\phi=\pi/4$ and it is equal to  ${\cal R} \cdot \Delta /2$. The trajectory of the photocentre is a straight line with stellar disk centre located in the middle of this line so that the peak to peak amplitude of the photocentre's displacement is twice this value. 

The angular radius of a star with solar radius located at 10 pc from the observer is ca. 500 $\mu$as. We can now connect the amplitude of the brightness variations, ${\cal A}_{\rm brightness}$, and astrometric jitter, ${\cal A}_{\rm jitter}$, for such a star: 
\begin{equation}
    {\cal A}_{\rm jitter} \, \textrm{[in } \mu \textrm{as]}    =5 \cdot {\cal A}_{\rm brightness}  \, \textrm{[in } \% \textrm{]}.
\label{eq:est}
\end{equation}

Over the last 40 years there have been several episodes when the transit of a large sunspot group caused a drop of the solar brightness by 0.2--0.3\% \citep[see, e.g., Fig.~2 from][]{Greg2016}. Equation~\ref{eq:est} then points to the amplitude of the astrometric jitter of 1--1.5 $\mu$as. Since such episodes of high variability are  rather rare, a more robust measure of solar and stellar rotational brightness variability might be the $R_{\rm var}$ metric introduced by \cite{Basri2010, Basri2011} for quantifying stellar brightness variations observed by the {\it Kepler} telescope. To calculate $R_{\rm var}$ first the peak-to-peak (or more precisely relative difference between the 95th and 5th percentile of the sorted flux values) variability in each of the 90-day {\it Kepler} quarters is calculated and then the median value among all quarters is taken.   The total duration of {\it Kepler} observations is about 4 years, so that $R_{\rm var}$ value corresponds to the mean variability over this period. By ``keplerizing'' available solar data \citep[see, e.g.,][]{basrietal2013},  \cite{Timo2020} showed that the mean value of the solar $R_{\rm var}$ over the last 140 years is 0.07\%, while the maximum value is 0.18\%. These numbers transfer to 0.35 and 0.9 $\mu$as amplitudes of the displacement, respectively. At the same time, \cite{Timo2020} showed that $R_{\rm var}$ values for stars with near-solar 
fundamental parameters and rotation periods can reach up to 0.7\% which would correspond to a displacement values of 3.5 $\mu$as. The $R_{\rm var}$ values of faster rotating G-dwarfs can reach up to 3\% \citep[see, Fig.~4 from][]{McQuillan2013} corresponding to a displacement of 15 $\mu$as.

The along-scan single-epoch precision of the Gaia measurements for the brightest stars is about 34  $\mu$as \citep[although a more optimistic value of 11  $\mu$as has also been used in the literature, see][for a discussion]{Perryman2014}. Consequently, our simple estimate shows that if the Sun were observed by Gaia from a distance of 10 pc, its activity would be too low to affect the measurements. At the same time our estimate shows that the astrometric jitter might become an important factor for more active stars especially since Gaia performs multiple measurements of the same stars \citep[with their number reaching up to ca. 100 for stars at intermediate galactic latitudes][]{Perryman2014}.

In the next sections of this paper we present much more accurate calculations of solar astrometric jitter based on the observed distribution of solar magnetic features and the SATIRE model. We expect that jitter returned by this model would reflect the jitter from other Sun-like stars with a similar activity level (if observed roughly equator-on).  In the subsequent publications we will present an extension to calculating astrometric jitter for stars with different fundamental parameters and with a broad range of magnetic activity values and observed at random  inclinations, i.e. angles between the direction to observer and rotation axis (see discussion in Sect.~\ref{sect:final}).
}

\section{Model description}\label{sect:model}
We utilize a highly precise version of the model, SATIRE-S, based on the satellite measurements of the magnetic field distribution on the solar disk \citep{SATIRE}. SATIRE-S  irradiance time series have been demonstrated to be consistent with solar observations from multiple sources 
\citep[see, e.g.,][and references therein]{balletal2014,yeoetal2014,Yeoetal2015,Danilovic2016}. The SATIRE-S model accounts for the irradiance variations brought about by magnetic features (namely bright faculae and dark sunspot umbrae and penumbrae) on the solar surface.  It uses full-disc magnetograms and intensity images of the Sun taken with daily cadence to determine the coverage of the solar disk by magnetic features. The intensities of the quiet Sun and magnetic features have been calculated by \cite{Unruhetal1999} with the ATLAS9 code \citep{kurucz1992,ATLAS9_CK}.

\subsection{{Photometric signal}}
The calculations of the solar  irradiance at any given day are done in the following steps. First, fractional coverages of each pixel of the magnetogram by magnetic features and quiet Sun (i.e. regions of the solar surface covered by neither faculae, nor spots) are calculated \citep[see][for details of the calculations]{yeoetal2014}. Second, for each magnetogram pixel the radiance from the region on the solar surface  corresponding to this pixel is computed. This is done by summing up radiance values from the quiet Sun and magnetic features, weighted with corresponding fractional coverages of the pixel. Finally, the irradiance from the full solar disk is calculated by summing up the contributions from all pixels.

Let us now apply such an approach for calculating solar irradiance passing through a filter with a given passband. The ratio between the magnetic component of the irradiance and the irradiance from the full-disk quiet Sun (i.e. Sun free from any magnetic features) can be found with the following equation:
\begin{equation}
{\cal B}(t) = \sum\limits_i \sum\limits_{k>0} \alpha_i^k(t) \left ( F_i^k-F_i^0 \right ) / \sum\limits_i F_i^0.
\label{eq:phot}
\end{equation}
Here the first summation (over $i$) is done over all solar disk pixels. The second summation (over $k$) is done over the three classes of magnetic features considered in the SATIRE-S model, i.e. faculae, spot umbrae, and spot penumbrae (corresponding to $k=1,\, 2,\, 3$, respectively). $\alpha_i^k(t)$ is the fractional coverage of the $i$-th pixel by the $k$-th component and $F_i^k$ is irradiance from the $i$-th pixel fully covered by the $k$-th component. $k=0$ corresponds to the quiet solar regions. 

The  $F_i^k$ values can be written as
{\begin{equation}
F_i^k=\Delta \Omega \int\limits_{\lambda} I (\lambda, \mu_i)\, {\cal \phi} (\lambda) \, d \lambda,   
\label{eq:flux}
\end{equation}
for detectors measuring energies (like all TSI instruments) and
\begin{equation}
F_i^k=\Delta \Omega \int\limits_{\lambda} I (\lambda, \mu_i)\, {\cal \phi} (\lambda)  / (h c / \lambda) \, d \lambda,   
\label{eq:photon}
\end{equation}
for detectors counting photons \citep[see, e.g. discussion in][]{Maxted2018}.}  $\Delta \Omega$ in Eq.~\ref{eq:flux}--\ref{eq:photon} is the solid angle of the region on the solar surface that corresponds to one pixel when the magnetograms  are obtained at 1 AU from the Sun. $I (\lambda, \mu_i)$ is the intensity from the $k$-th component. The intensities entering the SATIRE-S model can be written as a function of wavelength $\lambda$ and  the
cosine of the angle between the direction to the observer and the local stellar radius, $\mu_i$, corresponding to the $i$-th pixel (neglecting the change of the cosine value within the pixel). Finally, ${\cal \phi} (\lambda)$ is the transmission curve. {We used Eq.~\ref{eq:photon} for calculating solar astrometric and photometric signals as they would be measured by the photon-counting CCD detectors of Gaia and Small-JASMINE. Eq.~\ref{eq:flux} was used for the TSI calculations. Naturally, the radiometers used for the TSI measurements \citep[see, e.g.,][]{Greg2016} cannot be used for measuring stellar positions. Nevertheless, we also show astrometric signal as it would be measured in TSI to better illustrate its wavelength dependence.}

\subsection{{Astrometric signal}}
The shift of solar photocentre due to the magnetic activity can be found with an equation similar to Eq.~\ref{eq:phot}:
\begin{equation}
\begin{pmatrix} \bar X \\ \bar Y  \end{pmatrix} = \frac{\sum\limits_i \sum\limits_{k>0} \alpha_i^k(t)  \begin{pmatrix} X_i \\ Y_i  \end{pmatrix} \left (F_i^k -F_i^0     \right ) }{ \left ( 1+{\cal B} (t) \right) \sum\limits_i F_i^0 }. \label{eq:shift}   
\end{equation}
Here $\bar X$ and $\bar Y$ are shifts along and perpendicular to the equatorial plane, respectively. $X_i$ and $Y_i$ are the coordinates of the $i$-th pixel. The origin of the coordinate system is chosen to coincide with the solar disk centre.

In this study we calculate the trajectory of the solar photocentre with a daily cadence for the period from February 2, 1999 till August 1, 2014. The $\alpha_i^k$ values have been taken from \cite{yeoetal2014} who determined them based on the data from the Michelson Doppler Imager onboard the Solar and Heliospheric Observatory \citep[SOHO/MDI,][]{scherreretal1995} prior to April 30, 2010 and from the Helioseismic and Magnetic Imager onboard the Solar Dynamics Observatory \citep[SDO/HMI,][]{HMI, HMI_2} for the subsequent period. A special homogenising procedure has been applied to ensure the consistency between the two resulting segments of the reconstruction \citep[see][for more details]{yeoetal2014}.

\begin{figure}
\resizebox{\hsize}{!}{\includegraphics{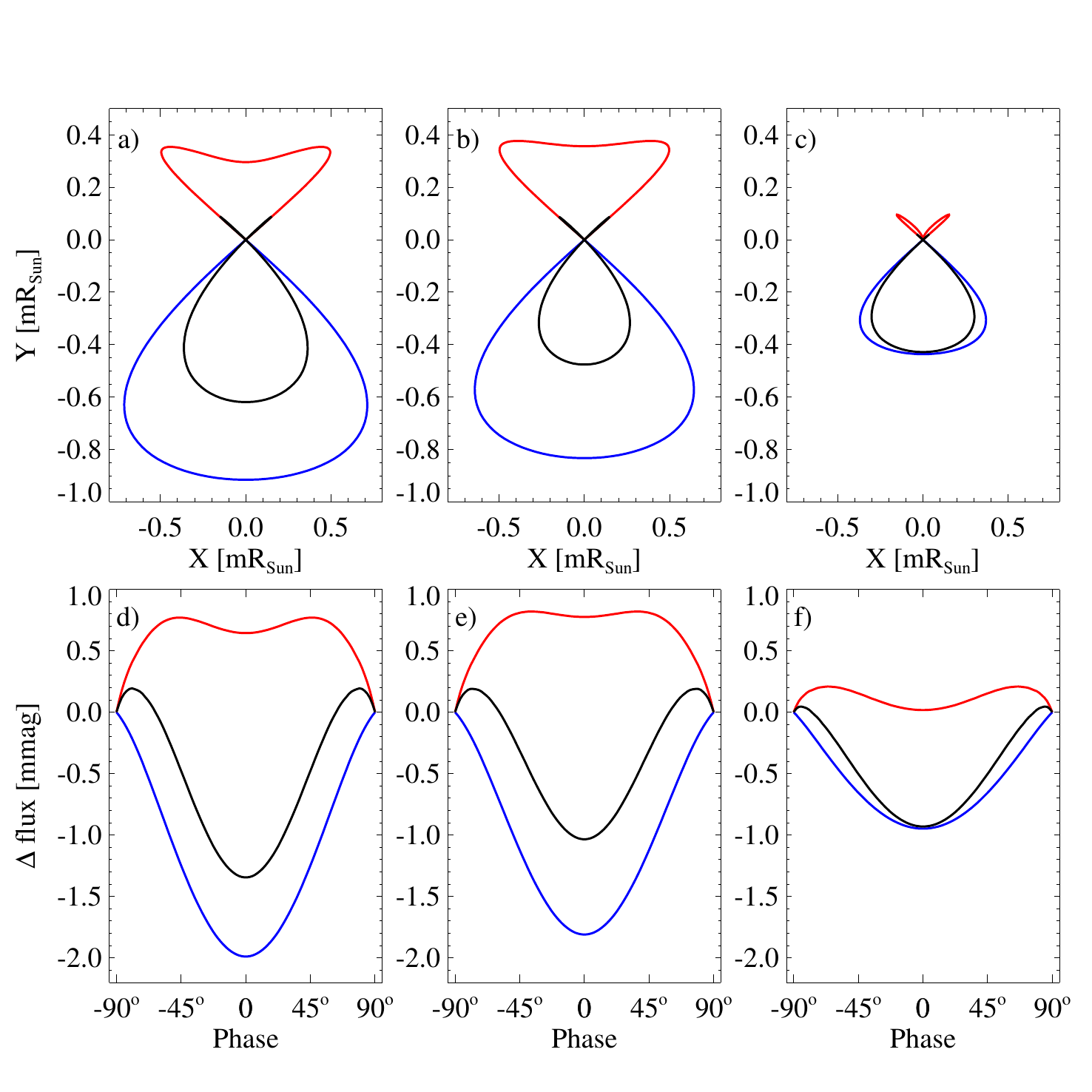}}
\caption{The daily displacements of the solar photocentre (top panels) and radiative flux changes (bottom panels) resulting from the transit of an active region (consisting of spot and facular parts, see detailed description in the text) plotted vs. phase of the transit.  Shown are total displacements and flux changes (black) as well as their facular (red) and spot (blue) parts. Calculations are performed in Gaia G passband (panels a and d), TSI (panels b and e), and  Small-JASMINE (panels e and f).}
\label{fig:model}
\end{figure}

\section{Results}\label{sect:res}
To better illustrate the origin of the solar photocentre displacement we start with considering a case of the hypothetical transit of a single active region across the visible solar disk before moving to a more sophisticated calculations of the displacements corresponding to the observed distribution of active regions. Namely, we consider the case of a relatively large active region \citep[see, e.g.,][for the size distribution of active regions]{BaumannSolanki2005} with a spot area of 3000 MSH (micro solar hemisphere) and a facular area of 9000 MSH {(corresponding to 0.6\% and 1.8\% coverages of the visible solar disk when observed near the disk centre, respectively)}. The ratio between the areas of the facular and spot parts of active region has been taken from \cite{Shapiro2020}.  Following \cite{wenzleretl2006} the spot area of the active region was set to consist of 80\% penumbra and 20\% umbra. We have considered the  non-evolving point-like (i.e. described by a single $\mu$ value) active region which transit solar disk at  the latitude of $30^{\circ}$.

Figure~\ref{fig:model} presents the displacements of the solar photocentre relative to the solar disk centre measured in milli solar radii ($\rm mR_{Sun}$) and radiative flux changes (measured in mmag) associated with such a transit. The displacements and brightness changes are plotted vs. the phase of the transit, which is, hereafter, defined as the longitude of the active region measured relative to the plane containing the observer and solar rotation axis. It equals to $-90^{\circ}$ when the active region just rotates in at the western limb, changes sign from minus to plus when the active region passes above the solar disk centre, and it equals to $90^{\circ}$ when the active region disappears at the eastern limb. We show the displacement and radiative flux changes as they would be measured in the Gaia G passband, the Total Spectral Irradiance (TSI, i.e. spectrally integrated radiative flux), and Small-JASMINE passband. The spectral transmission of the Gaia G passband has been taken from the Gaia DR2 revised passbands webpage\footnote{\url{https://www.cosmos.esa.int/web/gaia/iow_20180316}}. The G passband covers the wavelengths between 330 and 1050 nm \citep{Gaia_phot}. The exact spectral transmission of the Small-JASMINE mission \citep{JASMINE} is not yet available but it is anticipated that Small-JASMINE will make measurements in the 1.1--1.7 ${\rm \mu m}$ spectral interval. We have, therefore, considered a rectangular spectral transmission profile returning unity within this spectral interval and zero outside of it.

\begin{figure}
\resizebox{\hsize}{!}{\includegraphics{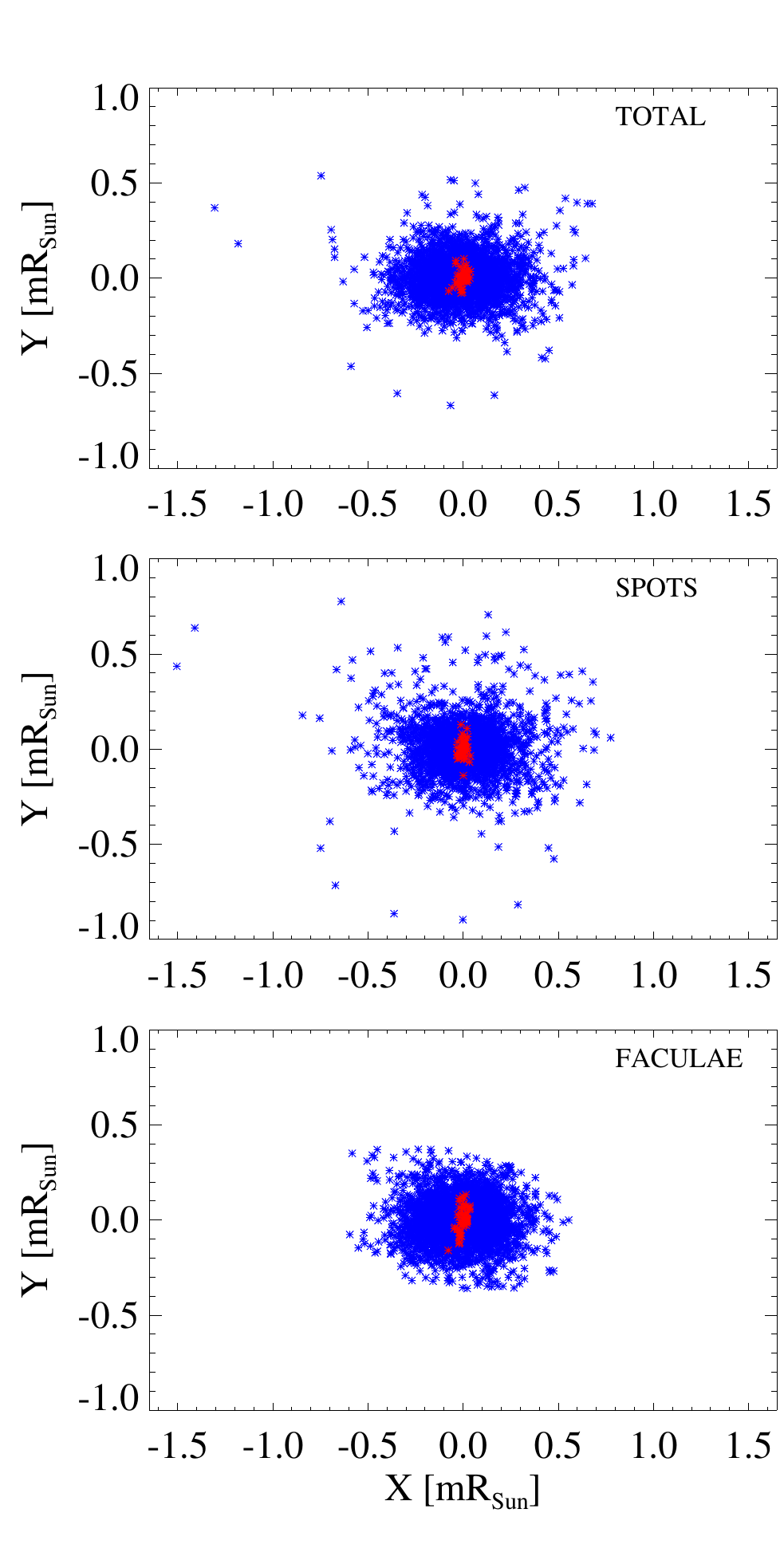}}
\caption{The daily (blue asterisks) and averaged over 81 day (red asterisks) displacements of the solar photocentre as they would be seen in the Gaia G passband for the period February 2, 1999 -- August 1, 2014. Shown are total displacements (top panel), as well as their spot (middle panel) and facular (bottom panel) components. X and Y axes are chosen to be along and perpendicular to the solar equator, respectively, and have equal scaling to better illustrate the shape of the photocentre trajectory. }
\label{fig:all}
\end{figure}

\begin{figure}
\resizebox{\hsize}{!}{\includegraphics{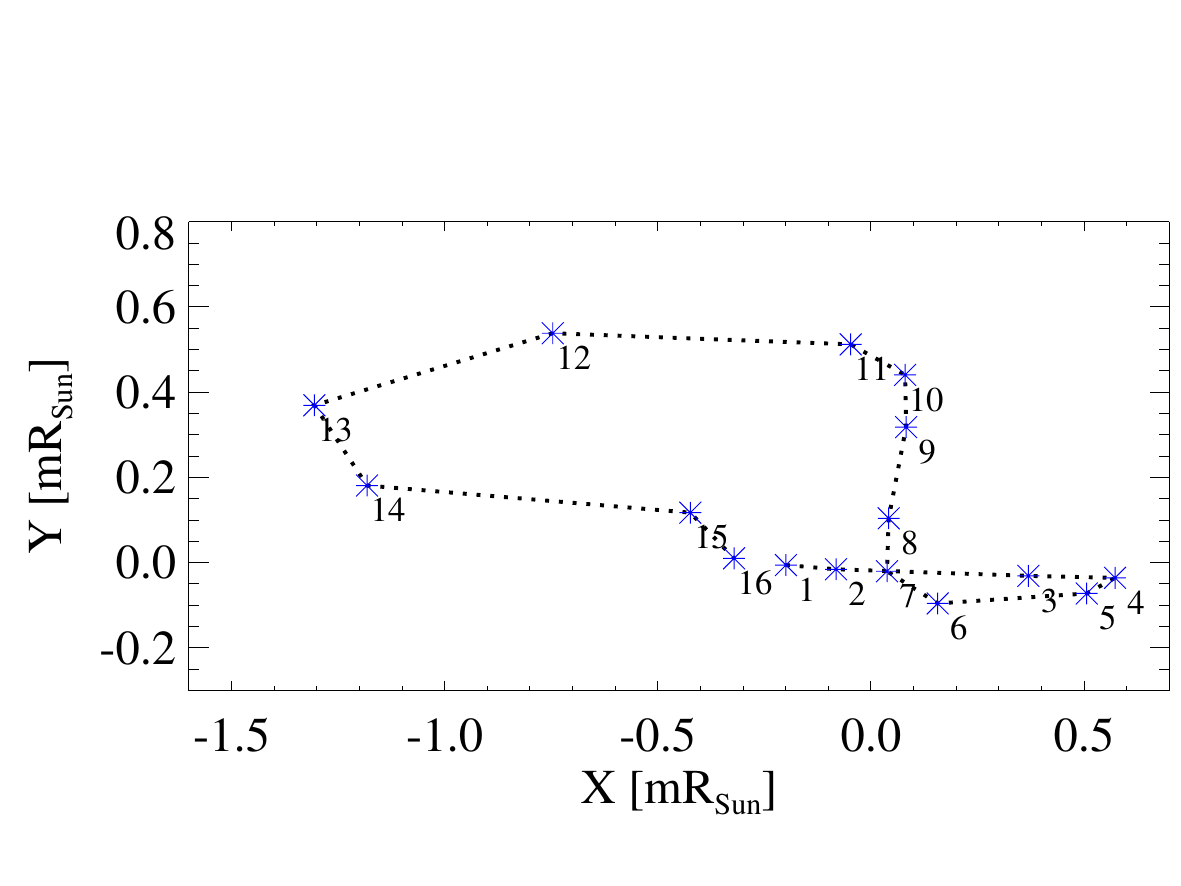}}
\caption{The daily displacements (blue asterisks) of the solar photocentre as they would be seen in the Gaia G passband for the period October 18, 2003 (day 1) -- November 3, 2003 (day 16).  The days are numbered from ``1''  to  ``16''.}
\label{fig:16}
\end{figure}
We put the origin of the coordinate system into the centre of the solar disk (which is assumed to be circular). The solar photocentre is located there before the active region rotates in. As soon as active region appears at the western part of the solar disk its dark spot part stats to repel the photocentre while its bright facular part starts to attract it. This is illustrated by the colour curves in Fig.~\ref{fig:model} which show how the trajectories of the solar photocentre would look if only spot and 
facular part (blue and red curves, respectively) of the active region transited the solar disk. One can see that the  transit of a  spot 
would cause the photocentre to move clockwise in the loop trajectory at negative ordinates and lead to the decrease of the solar brightness. At the same time the transit of a facular region would cause the photocentre to move clockwise  at positive ordinates  and leads to the increase of the solar brightness. The  facular brightness contrast noticeably decreases when facular region passes close to the disk centre (i.e. when it corresponds to near-zero phase values, see Fig.~\ref{fig:model}) leading to the decrease of the facular effect on the position of the photocentre and solar brightness. As a result, the transit of facular region leads to the heart-like shaped trajectory of the photocentre   and two-peak profile of the brightness change. Such a decrease of the facular contrast is especially pronounced for the Small-JASMINE passband (see Figs.~\ref{fig:model}cf) where it drops almost  to zero when active region crosses Y-axis.

The net effect of the transit of the active region is given by the competition between spot and facular effects. Since brightness contrast of faculae strongly increases towards the limb the facular effect overweights that of spot when active region is close to the limb. Consequently, after the active region rotates in at the western limb the photocentre first starts to move towards it, i.e. its  abscissa gets negative while the ordinate gets positive. This is clearly visible in all three trajectories (black curves in Figs.~\ref{fig:model}abc), although since facular contrast is decreasing in the infrared (see also below) the initial shift towards the active region is relatively small for the Small-JASMINE.

By the same token the active region first leads to the increase of the solar brightness. As active region rotates further from the limb the relative contribution of faculae becomes weaker and solar photocentre starts to move back towards the disk centre.  The photocentre follows a very similar trajectory it made moving towards the active region, e.g. these back and forth trajectories are indistinguishable in Fig.~\ref{fig:model}. After passing through the disk centre the solar photocentre makes a loop trajectory with negative ordinates before starting to move towards the active region again (which happens when the active region gets close to the eastern limb). Likewise, the effect of the active region on solar brightness remains negative (i.e. active region decreases the brightness) most of the time before increasing closer to the end of the transit. Naturally, after the transit (i.e. when active region rotates out of the visible solar disk ) the solar photocentre returns back to the disk centre and the solar brightness becomes equal to that of the quiet Sun.  

Figure~\ref{fig:model} demonstrates that the largest displacements are observed in Gaia G passband, which appears to be more appropriate for detecting stellar jitter than TSI (i.e. white light). One can see that the partial compensation between spot and facular components noticeably decreases the amplitude of the displacements and brightness changes in the Gaia G passband and TSI. The compensation is particularly strong in the TSI leading to the net effect of the active region transit there being  very similar to that  in the Small-JASMINE passband (for both displacement of the photocentre and brightness change), despite individual spot and facular effects are significantly smaller in the Small-JASMINE passband. This is because spot and facular contrasts both decrease towards the infrared but the facular contrast decreases faster. Consequently, the displacement of the photocentre and brightness changes in the Small-JASMINE passband are mainly brought about by the spot part of the active region and facular part has only a very small contribution to the photocentre displacement and brightness change.

Now we move to considering displacements caused by the observed distributions of magnetic features on the solar disk. Fig.~\ref{fig:all} shows the displacements in the Gaia G passband  over the whole period considered in this study. 
One can see that the clouds of points representing daily displacements (blue asterisks in Fig.~\ref{fig:all}) are  elongated along the solar equator (i.e. along the X-axis). This is because solar magnetic features are mainly concentrated in the low latitude regions of the Sun. For example, large spots rarely appear at latitudes higher than 30$^{\circ}$ and large facular regions are mainly concentrated at latitudes below 40$^{\circ}$. 

The displacements along the equator almost fully averages out when 81-day averaging is applied (red symbols in Fig.~\ref{fig:all}). This is because the activity in the western and eastern hemispheres of the Sun is on average the same. A larger averaged displacement remains along the Y-axis, which is caused by the 7.25$^\circ$ angle between the solar equator and ecliptic and the resulting apparent asymmetry of active region distributions along the Y-axis. We note that due to the Earth's orbit  around the Sun, the solar inclination (i.e. the angle between the rotation axis and direction to the observer) oscillates between 82.75$^\circ$ (in September and March)  and 90$^\circ$ (in December and June). Consequently, the solar photocentre oscillated with an annual period which is just a feature of the Sun observed from the vantage point of the Earth. Such an oscillation  will not be present in observations of other stars \citep[see also discussion in][]{Makarov2009} .

\begin{figure*}
\resizebox{\hsize}{!}{\includegraphics{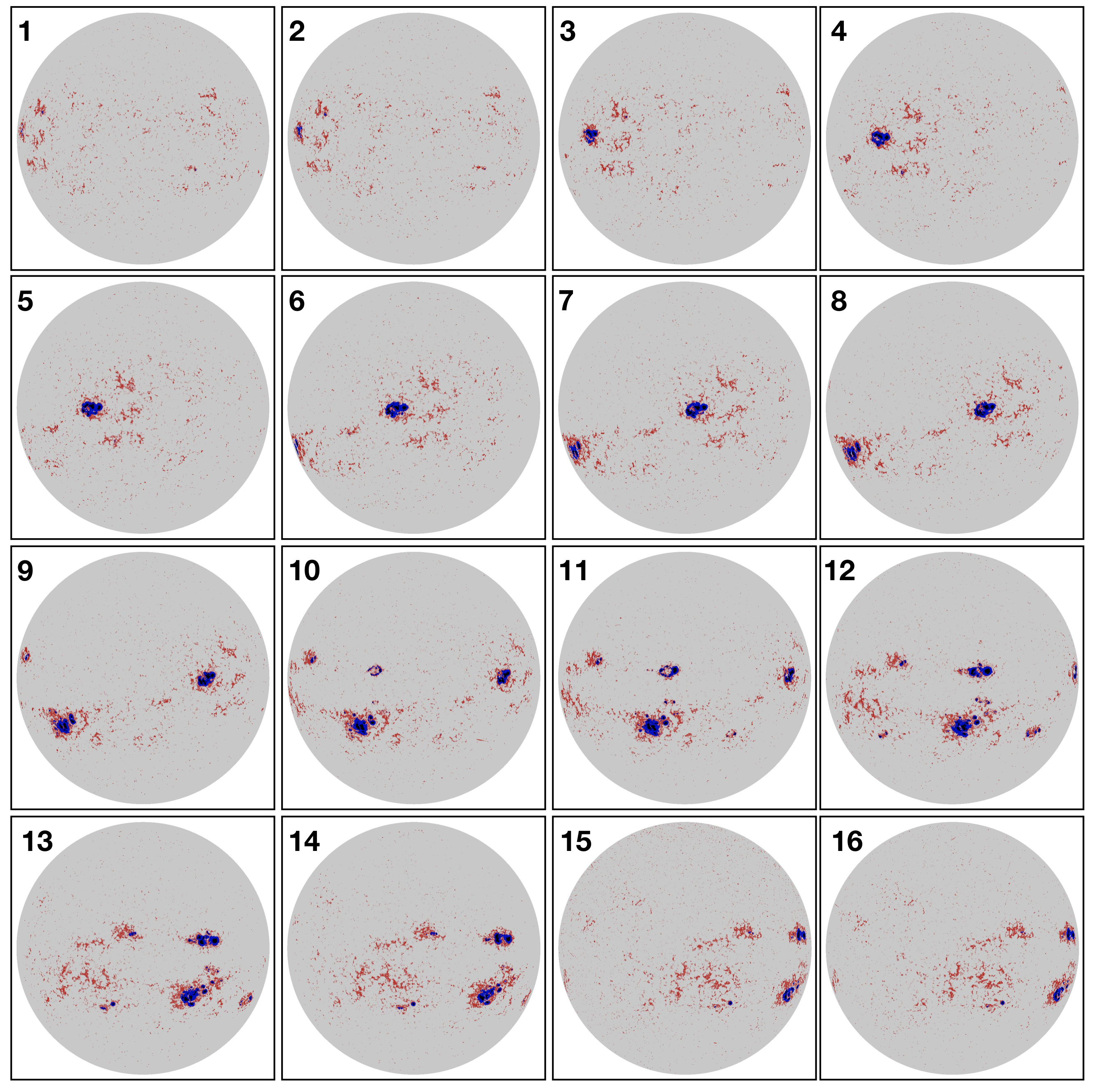}}
\caption{The surface distribution of faculae (red), penumbra (blue), and umbra (black) on the solar disk as it was observed for the period October 18, 2003 -- November 3, 2003 about midnight UTC at each day. The numbering of the days (in the left upper corner of each panel) corresponds to that in Fig.~\ref{fig:16}.}
\label{fig:images}
\end{figure*}

\begin{figure}
\resizebox{\hsize}{!}{\includegraphics{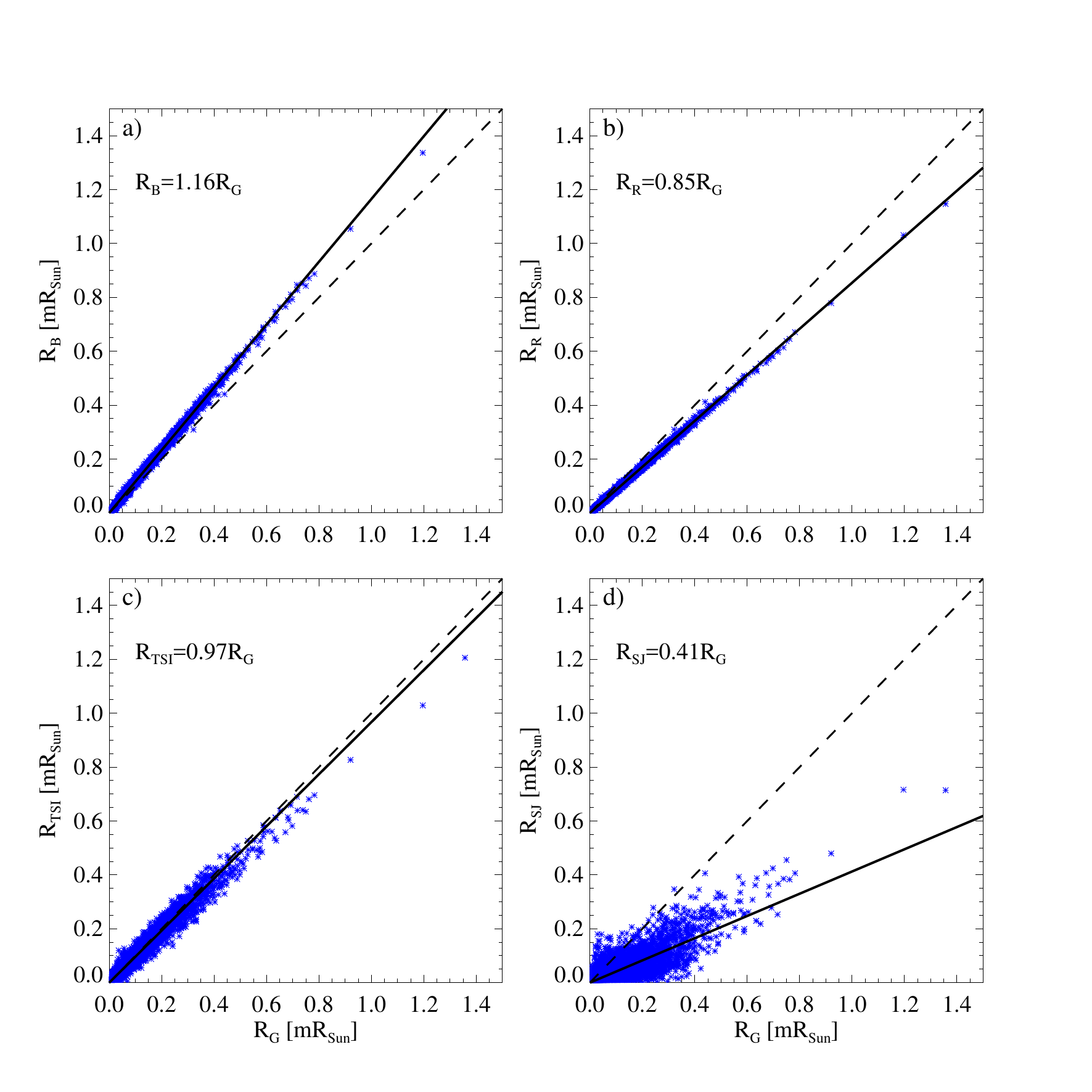}}
\caption{The displacements (relative to the solar disk centre) calculated in Blue (panel a) and Red (panel b) Gaia passbands as well as in the TSI (panel c) and in the Small-JASMINE passband (panel d) as functions of displacements calculated in the Gaia G passband. Each blue asterisk points to the daily values of the corresponding displacements. Solid black lines indicate the linear regression to the plotted dependences. Dashed lines represent identity lines.}
\label{fig:pass}
\end{figure}


The middle and bottom panels of Fig~\ref{fig:all} present displacements of the solar photocentre brought about by spots and facular features alone. One can see that, while the facular component of the displacement is relatively small and rather compact, the spot component has a number of larger excursions caused by transits of large spot groups. Since spot groups are accompanied by facular features which partly compensate displacements brought about by spots the excursions are more pronounced in the spot component than it the total displacement.

In Fig.~\ref{fig:16} we plot the trajectory of the solar photocentre during the period from October 18, 2003 till November 3, 2003  when three large active regions, consisting of spots and faculae  transited the solar disc. Daily snapshots of the  distributions of magnetic features on the solar surface are shown in Fig.~\ref{fig:images}. On October 18, 2003 (day 1 in Figs.~\ref{fig:16} and \ref{fig:images}) a large active region  just rotated onto the western part of the solar disk. As discussed previously, the faculae of the active region are especially bright when observed close to the solar limb. Consequently, on day 1 faculae exceed  the effect from spots causing  shift of the solar photocentre towards the active region, i.e. to the  west (negative abscissa values). As the active region rotates towards the disc centre the facular brightness contrast drops and at day 3 the effect of the spots surpasses  that of the faculae so that the solar photocentre starts to move away from the active region, shifting it  to the eastern part of the solar disk (positive abscissa values). Similarly, the active region rotating onto the disc on day 6 first attracts and then from day 7 starts to repel the photocentre. Since the active region is located at a relatively high  latitude the photocentre also experiences large displacements along the Y-axis. The trajectory of the photocentre is further affected by the sunspot group  emerged at day 10. Fig.~\ref{fig:images} shows that the area of the emerged group continues growing most of the time it is on the disc (till the effect of the foreshortening makes it less visible) amplifying the displacement of  the photocentre to the west. All in all,  both facular and  spot features affect the position of the photocentre. Furthermore, one can see that it is important to properly account not only for the solar rotation but also for the evolution of magnetic features to properly model the trajectory of the photocentre.
\begin{figure}
\resizebox{\hsize}{!}{\includegraphics{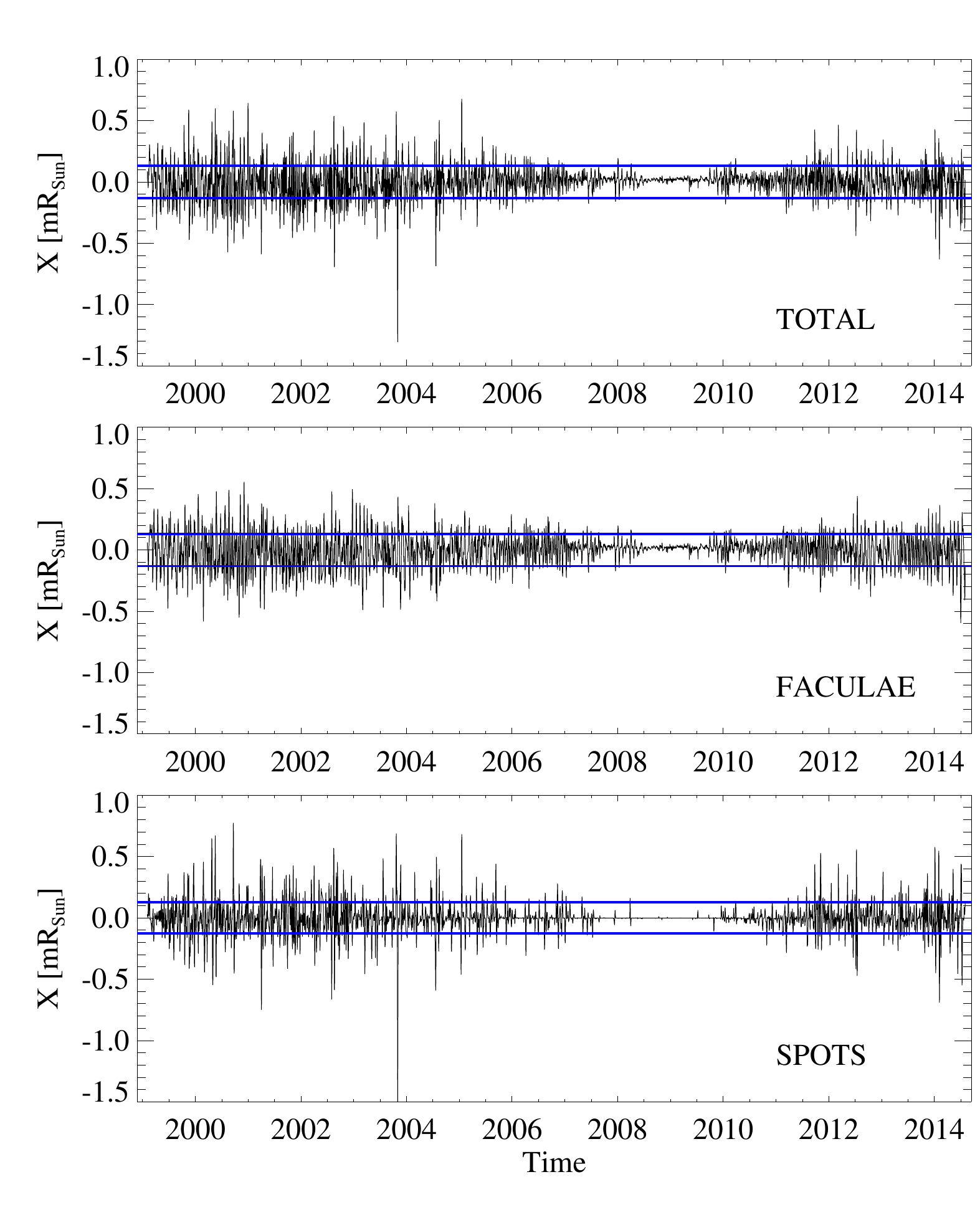}}
\caption{Displacements of the solar photocentre along the X-axis vs. time. Shown are total displacements (top panel) as well as their facular (middle panel) and spot (bottom panel) components. The displacements are calculated in the Gaia G passband. {Horizontal blue lines in each of the panels indicate corresponding values of the displacement standart deviation.}}
\label{fig:X}
\end{figure}

\begin{figure}
\resizebox{\hsize}{!}{\includegraphics{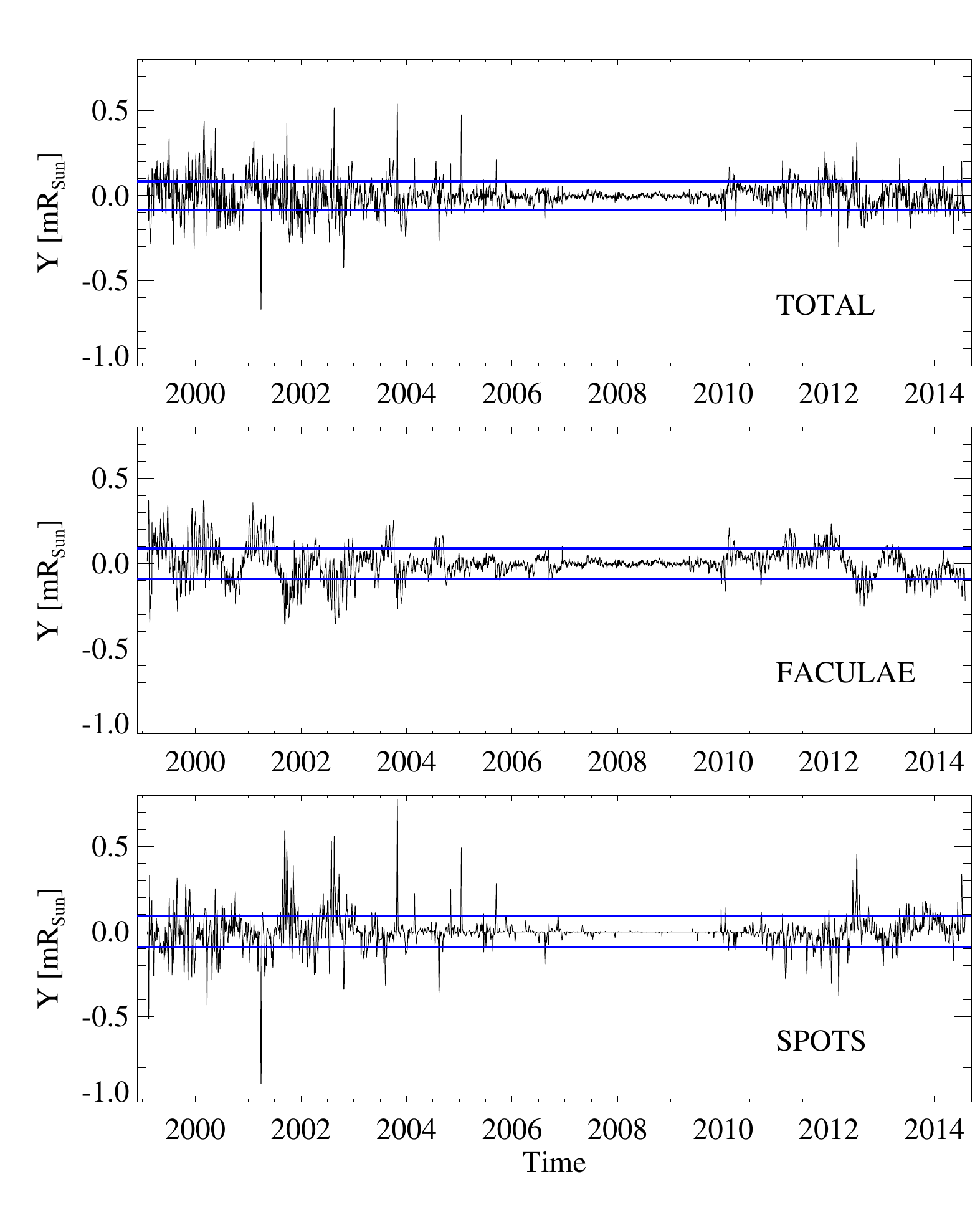}}
\caption{The same as Fig.~\ref{fig:X} but for the displacement along the Y-axis.}
\label{fig:Y}
\end{figure}

Displacements in Figs.~\ref{fig:all} and \ref{fig:16} have been calculated as they would be seen in the Gaia G filter. To better illustrate the dependence of the displacement on the spectral passband we also calculated the displacements as they would be seen in the Blue (330 -- 680 nm) and Red (630--1050 nm) Gaia photometers as well as in the TSI, and Small-JASMINE passband. Just like passband G, the transmission curves of the blue and red photometers have been taken from the Gaia DR2. We note that Gaia astrometric measurements are only available in the G passband so that displacements in the Red and Blue passbands {(similarly to the displacements in the TSI, see Sect.~\ref{sect:model})} are only shown to illustrate the effect of wavelength.

Figure~\ref{fig:pass} shows that displacements in the TSI and in all Gaia filters closely correlate with each other. The displacements in the TSI are quite close to those in the G filter (being on average only 3\% smaller, see right panel of Fig.~\ref{fig:pass}).  The largest displacement is found in the blue filter and the smallest in the red filter (16\% larger and 15\% smaller than in the G passbands, respectively). 

The behaviour of the displacements in the infrared Small-JASMINE passband is quite different: they show only  moderate correlation with  the displacements in the Gaia G filter  and also their amplitude is less than half that in the Gaia G filter.  Interestingly, Figs.~\ref{fig:pass}cd indicate that the displacements in the Small-JASMINE passband appear to be significantly smaller than those in the TSI. This implies that the effect of the compensation between facular and spot components of the displacements which caused TSI and  Small-JASMINE trajectories to look very similar in Figs.~\ref{fig:model}bc is smaller for the real distribution of solar magnetic features than for the case of the transit of single non-evolving active region. This can be explained, e.g. by considering displacements caused by the transits of active regions consisting of only faculae (since spot part of  active regions decays much faster than facular there are many such active regions on the Sun). Such facular active region would lead to a significant displacement of the solar photocentre in the TSI but due to a small facular infrared contrast will result in a very small displacement in the Small-JASMINE passband. One can, indeed, see that Fig.~\ref{fig:pass}d contains a lot of blue points close to the X-axis, i.e. points with noticeable displacement in the Gaia G passband but with very small displacement in the Small-JASMINE passband.


In Figs.~\ref{fig:X}~and~\ref{fig:Y} we plot the X- and Y-displacements in the Gaia G passband, respectively, as a function of time. One can see that that the amplitude of the jitter in the position of the solar photocentre is clearly modulated by solar activity, i.e. it is small during the activity minimum (2008--2009) and increases towards the maxima of cycles 23 and 24. As discussed above the conspicuous annual period in the Y-displacements is brought about by the change of the visible solar inclination caused by the Earth's orbital movement.

\section{Discussion and Conclusions}\label{sect:final}
We have extended the SATIRE model of solar irradiance variability to calculating the displacement of the solar photocentre caused by the magnetic activity of the Sun. {Such a displacement is caused by the dark spots and bright facular features on the solar surface and is often referred to as the astrometric jitter in the literature \citep[see, e.g.,][]{Makarov2010, Morris2018}.} We have calculated the displacements as they would be seen in the Gaia and Small-JASMINE passbands as well as in the TSI. The displacements are mainly visible on the timescale of the solar rotation and are caused by the transits of sunspot and facular features  over the visible solar disk as the Sun rotates. Our calculations indicate that facular and spot components of the displacement  have comparable amplitudes and thus the effect of faculae on the position of the photocentre cannot be neglected as has been previously done in a number of studies. 

The RMS amplitude of the displacements from the solar disk centre  as they would be seen in the Gaia G passband ($R_G(t)$, see Fig.~\ref{fig:pass}) during 2000 (a year of high solar activity) was about  0.24 $\rm mR_{Sun}$. For comparison, \cite{Makarov2010} found a somewhat smaller value of $0.91 \, \mu \rm AU \approx 0.19 \, mR_{Sun}$ for the same period. {One source of the difference between two estimates might be the fact that \cite{Makarov2010} used ground-based data for obtaining distribution of magnetic features on the Sun and for assessing their brightness contrasts.  At the same time, our calculations for 2000 rely on a distribution of magnetic features obtained from more accurate spaceborn data \citep[see detailed discussion in][]{yeoetal2014} and on calculated contrasts of magnetic features \citep{Unruhetal1999}.} Our model shows that the amplitude dropped to about 0.04 $\rm mR_{Sun}$  for  2008 (a year of low solar activity). At the same time individual peaks of the displacement along the solar equator often exceed  0.5 $\rm mR_{Sun}$ (see Fig.~\ref{fig:X}) with the most notable peak being in  November 2003, when a large sunspot group caused an excursion of almost 1.5  $\rm mR_{Sun}$ along the solar equator. Consequently, the amplitude of the solar astrometric jitter in the Gaia G passband is comparable to the signal caused by the Earth rotating around the Sun (about 0.6 $\rm mR_{Sun}$) but it is significantly lower than the signal produced by the orbital motion of Jupiter (about 1060 $\rm mR_{Sun}$). The amplitude of the jitter in the Small-JASMINE passband is expected to be more than two times smaller than that in the Gaia G passband.

The peak to peak amplitude of the solar jitter in the Gaia G passband reaches roughly 1 $\rm mR_{Sun}$ which corresponds to 0.5 $\mu$as for the Sun located 10 pc away from the observer \citep[which agrees with the result of][see, e.g., their Fig.~1]{Lagrange2011}. Interestingly, this number appears to be consistent with an estimate of 0.35--0.9 $\mu$as given in Sect.~\ref{sect:estimate}.

{ This is by far smaller than 
 the along-scan single-epoch precision of the Gaia measurements for the brightest stars (34  $\mu$as). At the same time the magnetic activity-induced displacements are expected to increase for more variable stars and for main sequence stars with larger radii. In particular, a simple estimate performed in Sect.~\ref{sect:estimate} suggests that the amplitude of the photometric jitter can reach up to 15 $\mu$as  for the most variable G-stars. }

The calculations presented in this paper have been performed for the Sun as it would be seen from its equatorial plane. { Recently, \cite{Emre2018} developed a model for flux emergence and transport on stars more active than the Sun, while \cite{Nina2, Nina1} proposed a method for calculating the disk distribution of  magnetic features as it would be seen out of the ecliptic. Finally,   \cite{witzkeetal2018, Veronika2020} showed how the contrasts of magnetic features depend on the stellar fundamental parameters.  In the next papers we plan to follow up  on these studies and 
extend calculations of stellar jitter to a) the Sun observed from an arbitrary inclination; b) the stars with near-solar magnetic activities but various fundamental parameters; c) stars more active than the Sun.}

\acknowledgments
We thank Jeff Kuhn for motivating us to perform this study and Kok Leng Yeo for providing masks of the magnetic features based on the SOHO/MDI and SDO/HMI data. The research leading to this paper has received funding from the European Research Council under the European Union’s Horizon 2020 research and innovation program (grant agreement No. 715947). It also got financial support from the BK21 plus program through the National Research Foundation (NRF) funded by the Ministry of Education of Korea. We would like to thank the International Space Science Institute, Bern, for their support of science team 446 and the resulting helpful discussions.

\bibliographystyle{aasjournal}



\end{document}